\documentclass[twocolumn,showpacs,preprintnumbers,amsmath,amssymb]{revtex4-1}

\usepackage{graphicx}% Include figure files
\usepackage{dcolumn}% Align table columns on decimal point
\usepackage{bm}% bold math
\usepackage{multirow}
\usepackage{float}
\usepackage{pdfpages}

\makeatletter
\providecommand \@ifxundefined [1]{%
 \@ifx{#1\undefined}
}%
\providecommand \@ifnum [1]{%
 \ifnum #1\expandafter \@firstoftwo
 \else \expandafter \@secondoftwo
 \fi
}%
\providecommand \@ifx [1]{%
 \ifx #1\expandafter \@firstoftwo
 \else \expandafter \@secondoftwo
 \fi
}%
\providecommand \natexlab [1]{#1}%
\providecommand \bibnamefont  [1]{#1}%
\providecommand \bibfnamefont [1]{#1}%
\providecommand \citenamefont [1]{#1}%
\providecommand \href@noop [0]{\@secondoftwo}%
\providecommand \href [0]{\begingroup \@sanitize@url \@href}%
\providecommand \@href[1]{\@@startlink{#1}\@@href}%
\providecommand \@@href[1]{\endgroup#1\@@endlink}%
\providecommand \@sanitize@url [0]{\catcode `\\12\catcode `\$12\catcode
  `\&12\catcode `\#12\catcode `\^12\catcode `\_12\catcode `\%12\relax}%
\providecommand \@@startlink[1]{}%
\providecommand \@@endlink[0]{}%
\providecommand \url  [0]{\begingroup\@sanitize@url \@url }%
\providecommand \@url [1]{\endgroup\@href {#1}{\urlprefix }}%
\providecommand \urlprefix  [0]{URL }%
\providecommand \selectlanguage [0]{\@gobble}%
\providecommand \bibinfo  [0]{\@secondoftwo}%
\providecommand \bibfield  [0]{\@secondoftwo}%
\providecommand \BibitemOpen [0]{}%
\providecommand \BibitemShut  [1]{\csname bibitem#1\endcsname}%
\let\auto@bib@innerbib\@empty
\begin{document}
%\openup 2em
\preprint{}

\title{Modeling non-harmonic behavior of materials from experimental inelastic neutron scattering and thermal expansion measurements}% Force line breaks with \\
%\thanks{A footnote to the article title}%

\author{Dipanshu Bansal}
\email{bansald@ornl.gov}
\affiliation{Materials Science and Technology Division, Oak Ridge National Laboratory, Oak Ridge, Tennessee 37831, USA}
\author{Amjad Aref}
\affiliation{University at Buffalo -- State University of New York, New York, 14260, USA}
\author{Gary Dargush}
\affiliation{University at Buffalo -- State University of New York, New York, 14260, USA}
\author{Olivier Delaire}
\email{olivier.delaire@duke.edu}
\affiliation{Mechanical Engineering and Materials Science, Duke University, Durham, North Carolina 27708, USA }
\affiliation{Materials Science and Technology Division, Oak Ridge National Laboratory, Oak Ridge, Tennessee 37831, USA}

%\openup 2em

\textbf{Notice: This manuscript has been authored by
UT-Battelle, LLC under Contract No. DE-AC05-
00OR22725 with the U.S. Department of Energy. The
United States Government retains and the publisher, by
accepting the article for publication, acknowledges that
the United States Government retains a non-exclusive,
paid-up, irrevocable, world-wide license to publish or
reproduce the published form of this manuscript, or
allow others to do so, for United States Government
purposes. The Department of Energy will provide
public access to these results of federally sponsored research
in accordance with the DOE Public Access Plan
(http://energy.gov/downloads/doe-public-access-plan).\vspace{0in}}

%\pagebreak

\date{\today}% It is always \today, today,
             %  but any date may be explicitly specified

\begin{abstract}
%\openup 2em

Based on thermodynamic principles, we derive expressions quantifying the non-harmonic vibrational behavior of materials, which are rigorous yet easily evaluated from experimentally available data for the thermal expansion coefficient and the phonon density of states. These experimentally-derived quantities are valuable to benchmark first-principles theoretical predictions of harmonic and non-harmonic thermal behaviors using perturbation theory, {\it ab initio} molecular-dynamics, or Monte-Carlo simulations. We illustrate this analysis by computing the harmonic, dilational, and anharmonic contributions to the entropy, internal energy, and free energy of elemental aluminum and the ordered compound ${\rm FeSi}$ over  a wide range of temperature. Results agree well with previous data in the literature and provide an efficient approach to estimate anharmonic effects in materials.
\end{abstract}

%\pacs{Valid PACS appear here}% PACS, the Physics and Astronomy
                             % Classification Scheme.
%\keywords{Suggested keywords}%Use showkeys class option if keyword
                              %display desired
\maketitle

\section{Introduction}
When considering the thermal properties of a crystalline material based on an ensemble of quantum harmonic oscillators, it is customary to start by considering a fixed set of oscillator frequencies. For a crystalline solid, this corresponds to having a phonon density of state (DOS) that does not depend on temperature or volume (pressure). Within this assumption, the Bose-Einstein occupation factor, together with the quantization of energy levels, provides the link between the set of oscillator frequencies and the free energy, internal energy, and entropy of the system. However, thermal expansion and thermal resistivity cannot be reconciliated within the crude assumption of purely harmonic oscillators. In the next-level approximation of so-called quasi-harmonic oscillators, the oscillator frequencies are taken to depend on volume, which recovers a finite thermal expansion coefficient, as well as a partial account of the commonly observed rise of the lattice specific heat above the Dulong-Petit constant value of $3R$ at high temperatures ($R$ is the universal gas constant). Despite the convenience of this quasi-harmonic approximation (QHA), high-temperature thermal properties derived within QHA often remain inaccurate, and deviations from harmonic or quasi-harmonic behaviors need to be quantified. Phonon-phonon interactions leading to intrinsic temperature dependence (at constant $V$) of phonon frequencies, as well as finite thermal resistivity, are generally termed ``intrinsically anharmonic''  \cite{MottJones1958, Mermin, Kittel1966, Wallace1, Wallace2, Grimvall_book, Fultz2010}. In the present study, we follow the same terminology, but we note that this intrinsic anharmonicity associated with explicit temperature dependence of renormalized harmonic effective force-constants may arise from either non-quadratic terms in the interatomic potential (leading to phonon-phonon interactions and renormalization of the quadratic force-constants), or owing to the influence of other degrees-of-freedom (for example via adiabatic electron-phonon coupling or magnon-phonon coupling). In addition, the phonon quasiparticles acquire finite lifetimes owing to phonon-phonon, phonon-electron, or phonon-magnon interactions \cite{Wallace1, Wallace2, Mermin, Kittel1966, Grimvall_book, Fultz2010, Grimvall1991, Katsnelson2003, Wallace1992, Plakida1969, Ma2013, Delaire2011, Manley2014, Budai2014, Calder2015, Chen2015, Bansal_2015}, which provide various channels for annihilation and creation of the quasiparticles. In the present study, we will consider the effect of intrinsic anharmonicity on the temperature dependence of phonon spectra and the high-temperature thermodynamics, assuming that the quasiparticle picture of phonon oscillators remains valid, and we will not consider the effect of phonon lifetimes on equilibrium thermodynamics, which is likely to be a weak effect in most crystals.

Within the QHA, one can relate the volumetric thermal expansion coefficient $\alpha$ to the Gr\"uneisen parameter $\gamma$, specific heat capacity at constant volume $C_V$, and bulk modulus $B$ by $\gamma = \alpha C_V B / V_0$, where $V_0$ is the volume at the equilibrium atomic configuration. This relationship defines an average, or ``thermodynamic'' Gr\"uneisen parameter \cite{Mermin, Kittel1966, MottJones1958, Wallace1, Wallace2, Grimvall_book}. Alternatively, mode-specific Gr\"uneisen parameters can be obtained from the volume dependence of the frequency for the phonon mode $k, s$, as $\gamma_{ks} = - \partial \ln \omega_{ks} / \ln V$. This mode-dependent quantitiy may be obtained either experimentally through pressure-dependent phonon measurements (for example in diamond anvil cells) or computationally (for example via density-functional theory simulations of phonon dispersions at different volumes). Yet another approach considers an experimentally-evaluated ``thermal effective'' Gr\"uneisen parameters from the measured temperature dependence of phonon frequencies, for example with inelastic neutron scattering (INS), inelastic x-ray scattering (IXS), or optical spectroscopy measurements of phonon frequencies as a function of temperature. Often, the ``thermodynamic'' Gr\"uneisen parameter and  the ``thermal effective'' Gr\"uneisen parameters differ from each other, suggesting that the assumption of phonon frequencies depending on the volume alone cannot be reconciled with the experimentally observed phonon softening/stiffening. Thus, the explicit dependence of phonon frequencies on temperature reveals that a system has a more complex anharmonic behavior \cite{Wallace1, Wallace2, Grimvall_book, Mermin, Kittel1966, Grimvall1991, Katsnelson2003, Wallace1992, Plakida1969, Fultz2010}. 

In modern first-principles materials simulations, the Gr\"uneisen parameter is generally derived from phonon dispersions calculated with either perturbation theory, or the so-called ``direct method'' based on computing forces resulting from atomic displacements in a supercell \cite{Baroni1987, Gonze1995_1, Gonze1995_2, phonopy_1, Parlinski2000}. While increasingly accurate and manageable, these computational methods are often still based on harmonic or QHA treatments, and they also remain challenging to apply to complex unit cells and disordered alloys, or materials containing defects. These shortcomings thus motivate experimental confirmation. First-principles phonon simulations can also result in unphysical predictions of unstable phonon modes for strongly anharmonic systems, and in such cases their applicability to high-temperature non-harmonic effects can be severely limited. In such situations, ab-initio molecular dynamics (AIMD) simulations, including anharmonic effects at finite temperature (with quasi-classical Boltzmann phonon statistics), can be particularly valuable, despite their computational cost. In order to accurately capture anharmonic effects from first-principles while limiting computational requirements, a number of theoretical methods have recently been devised \cite{Souvatzis2008, Errea2013, Ljungberg2013, Zhang2014, Monserrat2013, Hellman2013, Glensk2015}, which will benefit from direct comparison with experimental data probing anharmonicity. To this end, we present expressions for thermodynamic functions describing the vibrational behavior of non-harmonic materials, which can be conveniently evaluated from experimental INS and thermal expansion measurements. As illustration, we have computed the harmonic, quasi-harmonic, anharmonic, and dilational vibrational contribution to the free energy, internal energy, and entropy of the lattice for elemental aluminum and FeSi. This practical procedure provides a convenient experimental estimate of anharmonic effects in materials.

%\begin{figure}
%\begin{center}
%\includegraphics[width=0.5\textwidth]{figures/Aluminum/Al_PDoS-eps-converted-to.pdf}% Here is how to import EPS art
%\caption[Phonon density of state data of aluminum at different temperatures obtained from experimental inelastic neutron scattering.]{\label{Al_PDoS}Phonon density of state data of aluminum at different temperatures obtained from experimental inelastic neutron scattering (adapted from \cite{Tang2010}).}
%\end{center}
%\end{figure}
%
%Figure~\ref{Al_PDoS} shows the experimental inelastic neutron scattering data of aluminum at different temperatures. 

\section{Derivation}\label{BHQA}

The total free energy of a non-magnetic lattice is comprised of mainly four parts \cite{Wallace1, Wallace2, Grimvall_book, MottJones1958, Grabowski2009},
\begin{equation}\label{ENS_pre1}
F = F_{el} + F_{vib,qh} + F_{vib,ah} + F_{vac},
\end{equation}
where $F_{el}$, $F_{vib,qh}$, $F_{vib,ah}$, and $F_{vac}$ represent the electronic, vibrational quasi-harmonic, vibrational anharmonic, and vacancy contribution to the free energy, respectively. The vibrational quasi-harmonic contribution to the free energy can be further decomposed into harmonic, $F_{vib,h}$ and dilational, $F_{vib,d}$ components. Thus:
\begin{equation}\label{ENS_pre2}
F_{vib} =  F_{vib,qh} + F_{vib,ah} = F_{vib,h} + F_{vib,d} + F_{vib,ah}.
\end{equation}
The derivative of the total free energy with respect to volume at constant temperature can be calculated to obtain pressure $P$ \cite{Wallace1, Wallace2, Grimvall_book, MottJones1958},
\begin{equation}\label{ENS_pre3}
\left(\frac{\partial F}{\partial V}\right)_T = -P = -P_{el} -P_{vib} - P_{vac}. 
\end{equation}
From Eq.~\eqref{ENS_pre3}, the coefficient of volumetric thermal expansion, $\alpha$, can be determined by \cite{Wallace1, Wallace2, Grimvall_book, MottJones1958},
\begin{align}\label{ENS_pre4}
\alpha = \chi\left(\frac{\partial P}{\partial T}\right)_V &= \chi\left(\frac{\partial(P_{el} + P_{vib} + P_{vac})}{\partial T}\right)_V  \nonumber \\
&= \alpha_{el} + \alpha_{vib} + \alpha_{vac},
\end{align}
where $\chi$ is the compressibility of the material at temperature $T$. In addition, the subscripts $el$, $vib$, and $vac$ in Eq.~\eqref{ENS_pre3}, and Eq.~\eqref{ENS_pre4} represent the electronic, vibrational, and vacancy contribution to the pressure and the thermal expansion coefficient, respectively. Subsequently, by subtracting the electronic and vacancy contributions from the total volumetric thermal expansion coefficient, we can relate the total vibrational free energy to the (vibrational) thermal expansion coefficient. 

In the harmonic/quasi-harmonic oscillator approximation, we can calculate the Hamiltonian for $3N$ independent oscillators, one for each wave vector $k$ and phonon branch index $s$. Thus, we can specify the eigenstate by giving a set of $3N$ quantum numbers. The energy of the quantized eigenstate is given by \cite{Mermin, Kittel1966, MottJones1958},
\begin{equation}\label{ENS1}
E_{ks} = \sum\limits_{k,s}(n_{ks} + \frac{1}{2})h\nu_{ks}
\end{equation}
where $n_{ks} = \frac{1}{e^{h\nu_{ks}/k_{\rm B}T}-1}$ is the Bose-Einstein distribution, and $\nu_{ks}$ is the characteristic frequency of  normal mode $(k,s)$ of the crystal, $k_{\rm B}$ is Boltzmann's constant, $T$ is temperature, and $h$ is Planck's constant.
Thus, knowing the energy of the eigenstate, the internal energy of a lattice can be written as,
\begin{equation}\label{ENS2}
U = U_{eq} + \frac{1}{2}\sum\limits_{ks}h\nu_{ks} + \sum\limits_{ks}\frac{h\nu_{ks}}{e^{h\nu_{ks}/k_{\rm B}T}-1}
\end{equation}
where $U_{eq}$ is the energy in the equilibrium configuration. For anharmonic oscillators, it is generally much harder to calculate the independent quantized energy eigenstates. In order to calculate the thermodynamics of an anharmonic crystal, whose vibrational behavior can still be described by a set of phonon-like quasiparticles, we infer the total internal and free energy of the system from the experimentally measured thermal expansion coefficient and phonon DOS. We note that in doing so, we will lose information related to the wave vector and branch index of different phonon modes. 

For non-interacting phonon-like quasiparticles, the vibrational entropy $S_{vib}$ under a grand canonical ensemble is given by \cite{Wallace1, Wallace2, Grimvall_book},
\begin{align}\label{ENS3}
S_{vib} = k_{\rm B}\sum\limits_{k,s}&\left[(1+n_{ks})\ln(1+n_{ks})\right. \nonumber\\ 
&\left.- n_{ks}\ln(n_{ks})\right]
\end{align}
Furthermore, using the phonon density of states $g(E)$, Eq.~\eqref{ENS3} is transformed to an integral,  
\begin{align}\label{ENS4}
S_{vib} = 3Nk_{\rm B}\int g(E)&\left[(1+n(E))\ln(1+n(E))\right. \nonumber\\
&\left. - n(E)\ln(n(E))\right]~dE
\end{align}
where $N$ is the number of atoms in a crystal. We should mention here that in the derivation of Eq.~\eqref{ENS3}, one typically uses the QHA for the quantized energy eigenstates. Nevertheless, as shown by Barron \cite{Barron1963},  Cochran and Cowley \cite{Cochran}, and Hui and Allen \cite{Hui1975}, the entropy is thought to still be accurately  accounted for by Eq.~\eqref{ENS3}, if one replaces the QHA vibrational frequencies with anharmonically renormalized phonon frequencies, as experimentally measured at finite temperature. Thus, we will proceed by evaluating Eq.~\eqref{ENS3} and Eq.~\eqref{ENS4} with the experimentally determined vibrational spectra at different temperatures. 

Considering $E$ the total internal energy of the system under consideration (assuming any anharmonic effects are already included), the total vibrational free energy is given by:
\begin{align}\label{ENS5}
F_{vib} &= E - TS_{vib}.
\end{align}
To this free energy we must add the energy of atoms in the solid at volume $V$. Accordingly, the total vibrational free energy is,
\begin{align}\label{ENS6}
F_{vib} &= N\epsilon + E - TS_{vib}. 
\end{align}
Now, the derivative of the vibrational free energy with respect to volume at a given lattice configuration, $A$, can be calculated at constant temperature, $T$, from Eq.~\eqref{ENS6} as follows:
\begin{align}\label{ENS7_1}
\left(\frac{\partial F_{vib}}{\partial V}\right)_T = \left(N\frac{\partial\epsilon}{\partial V} + \frac{\partial E}{\partial V}  - \frac{\partial(TS_{vib})}{\partial V} \right)_T^A = -P_{vib}.
\end{align}
We can also evaluate the derivative of the vibrational free energy with respect to volume at lattice configuration, $B$, infinitesimally away from $A$, and at constant $T$, and subtract the two to obtain the differential:
\begin{align}\label{ENS7}
&\delta\left(\frac{\partial F_{vib}}{\partial V}\right)_T = \left(N\frac{\partial\epsilon}{\partial V} + \frac{\partial E}{\partial V}  - \frac{\partial(TS_{vib})}{\partial V} \right)_T^B \nonumber \\
&\hspace{0.4in}- \left(N\frac{\partial\epsilon}{\partial V} + \frac{\partial E}{\partial V}  - \frac{\partial(TS_{vib})}{\partial V} \right)_T^A= -\delta (P_{vib}).
\end{align}
Furthermore, to evaluate the derivative, $\frac{\partial\epsilon}{\partial V}$, we need to calculate the work done by the system for an infinitesimal change of the lattice volume. The work required to change the atomic energy from $\epsilon_A$ to $\epsilon_B$ can be expressed as,
\begin{equation}\label{TENS7}
W  = N(\epsilon_B-\epsilon_A).
\end{equation}
The work required in expansion/contraction of the crystal from volume $V_A$ to volume $V_B$ can also be determined from the compressibility, $\chi$, as follows:
\begin{equation}\label{TENS8}
W = \int\limits_{V_A}^{V_B}\frac{V-V_A}{V}\frac{1}{\chi(T)}dV.
\end{equation}
We can now equate the work required for an infinitesimal change of the lattice volume from Eq.~\eqref{TENS7} and~\eqref{TENS8},
\begin{equation}\label{TENS9}
N(\epsilon_B-\epsilon_A) = \int\limits_{V_A}^{V_B}\frac{V-V_A}{V}\frac{1}{\chi(T)}dV.
\end{equation}
Subsequently, by differentiating Eq.~\eqref{TENS9} with respect to volume at constant temperature, we obtain
\begin{equation}\label{TENS10}
N\left(\frac{\partial\epsilon_B}{\partial V}-\frac{\partial\epsilon_A}{\partial V}\right)_T = \frac{V_B-V_A}{V_B}\frac{1}{\chi(T)},
\end{equation}
or, considering $B$ as any general lattice configuration with $V$ infinitesimally close to configuration $A$, we obtain,
\begin{align}\label{TENS11}
\left(N\frac{\partial\epsilon}{\partial V}\right)_T = \left(N\left.\frac{\partial\epsilon_A}{\partial V}\right|_{V=V_A} + \frac{V-V_A}{V}\frac{1}{\chi(T)}\right)_T \; .
\end{align}

Now, using Eq.~\eqref{TENS11}, we can re-write the differential change in volume derivative of the free energy in Eq.~\eqref{ENS7} as follows:
 \begin{align}\label{ENS13}
\delta\left(\frac{\partial F_{vib}}{\partial V}\right)_T &= \left(N\left.\frac{\partial\epsilon_A}{\partial V}\right|_{V=V_A} + \frac{V-V_A}{V}\frac{1}{\chi(T)}\right. \nonumber \\
&+ \left.\left(\frac{\partial E}{\partial V}  - \frac{\partial(TS_{vib})}{\partial V}\right)\right|_{V=V} -N\left.\frac{\partial\epsilon_A}{\partial V}\right|_{V=V_A} \nonumber \\
&-\left.\left.\left(\frac{\partial E}{\partial V}  - \frac{\partial(TS_{vib})}{\partial V}\right)\right|_{V=V_A}\right)_T =  -\delta (P_{vib}).
\end{align}
Furthermore, differentiating Eq.~\eqref{ENS13} with respect to temperature at constant pressure or with respect to temperature at constant volume, and using the expression for the volumetric thermal expansion coefficient, $\alpha_{vib}(T) = \frac{1}{V}\left(\frac{\partial V}{\partial T}\right)_P = \chi\left(\frac{\partial P_{vib}}{\partial T}\right)_V $, we can calculate
\begin{align}\label{ENS14}
\alpha_{vib}(T) &= \alpha_{vib}(T_A) -\chi(T_A)\nonumber \\
&\hspace{0.1in} \times\left[\left.\left(\frac{\partial^2 E}{\partial V\partial T}  - \frac{\partial^2(TS_{vib})}{\partial V\partial T}\right)\right|_{V=V}\right. \nonumber \\
&\hspace{0.1in}- \left.\left.\left(\frac{\partial^2 E}{\partial V\partial T}  - \frac{\partial^2(TS_{vib})}{\partial V\partial T}\right)\right|_{V=V_A}\right] \nonumber \\
&= \alpha_{vib}(T_A) -\chi(T_A)\times \delta\left(\frac{\partial^2 E}{\partial V\partial T}  - \frac{\partial^2(TS_{vib})}{\partial V\partial T}\right).
\end{align}
Lastly, we can numerically integrate the expression in Eq.~\eqref{ENS14} to calculate the total (including harmonic, dilational, and anharmonic contributions) vibrational internal energy $E$ and vibrational free energy $F_{vib}$ of the crystal from the available experimental data for thermal expansion and the phonon DOS. However, one has to be careful to add a proper integration constant, i.e., $S~dT$, when  integrating with respect to the volume. 

\begin{figure}
\begin{center}
\includegraphics[trim=29.0cm 0.5cm 5cm 0.5cm, clip=true, width=0.45\textwidth]{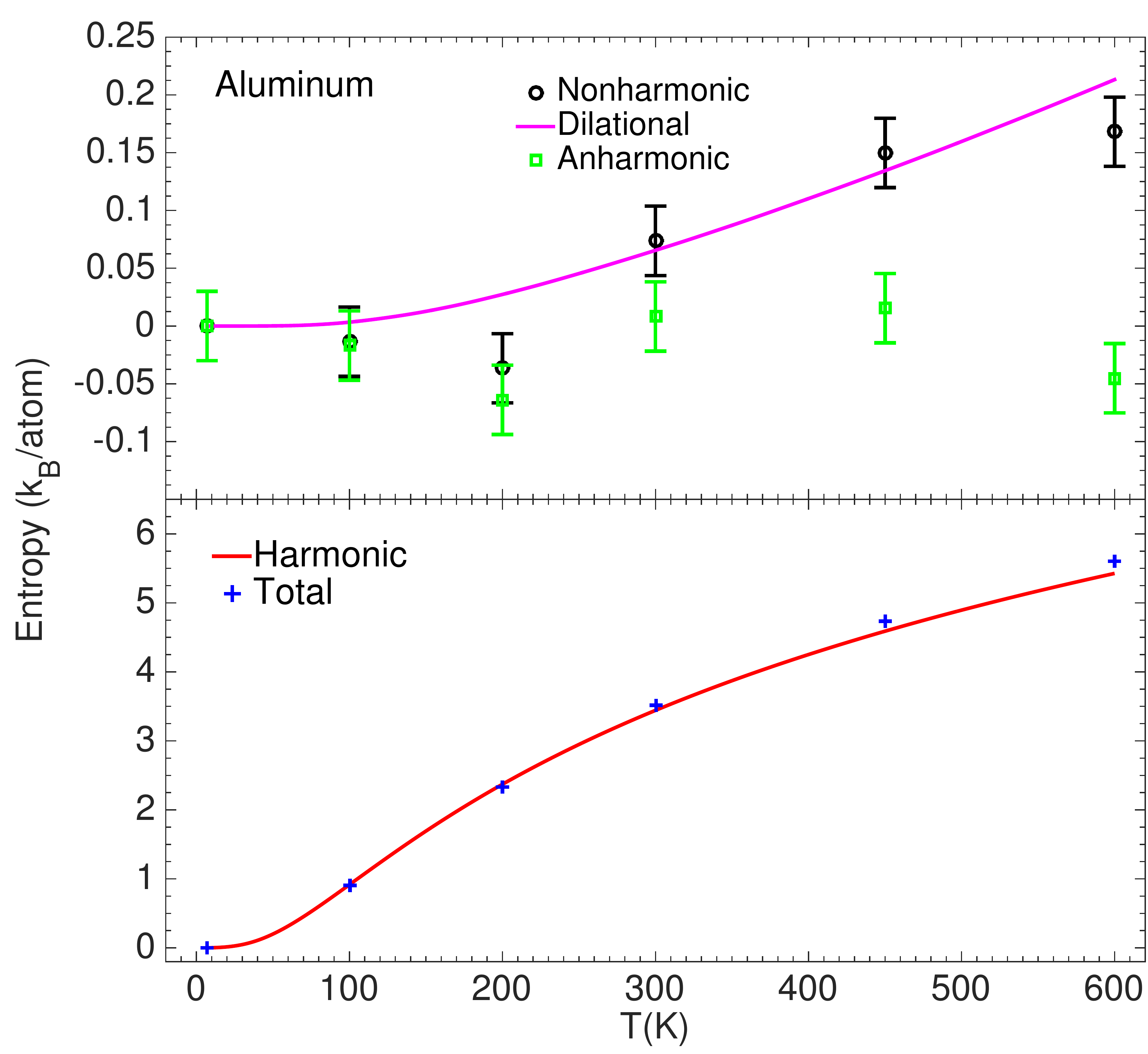}% Here is how to import EPS art
\caption{\label{entropy_Al}Variation of harmonic, nonharmonic, total, dilational, and anharmonic vibrational entropy $S = S(T) - S(T=0\,K)$ of aluminum crystal with temperature. The error in entropy values are estimated from the error bar in phonon density of states.}
\end{center}
\end{figure}

\section{Examples}
\subsection{Aluminum}
To evaluate the expression in Eq.~\eqref{ENS14} for the elemental metal aluminum, we used the experimental volumetric thermal expansion data from Wang and Reeber \cite{Wang2000}, and temperature dependent compressibility values from He \emph{et al.} \cite{He2004}. Grabowski \emph{et al.} \cite{Grabowski2009} have calculated the electronic and vacancy thermal expansion coefficient ($\alpha_{el}$ and $\alpha_{vac}$) from {\it ab initio} studies. The vibrational thermal expansion coefficient $\alpha_{vib}$ has been calculated by subtracting the electronic and vacancy contribution from the total thermal expansion coefficient. 

Figure~\ref{entropy_Al} shows the variation in harmonic, nonharmonic, total, dilational, and anharmonic vibrational entropy of aluminum crystal with temperature. The harmonic entropy $S_{vib,h}$ has been calculated using Eq.~\eqref{ENS3} with $g(E)$ measured at $T = 7$\,K, while the total entropy $S_{vib}$ has been calculated using Eq.~\eqref{ENS3} with $g(E)$ measured at different temperatures, as reported by Tang \emph{et al.} \cite{Tang2010}. The dilational entropy is evaluated as:
\begin{align}\label{ENS15}
S_{vib,d} &= \int\limits_0^T~\frac{(C_p-C_v)}{T'}~dT'  \nonumber \\
&= \int \limits_0^T~\frac{\alpha_{vib}(T')^2V(T')}{\chi(T')}~dT'
\end{align}
where $C_p$ and $C_v$ are the specific heat capacity of the material at constant pressure and temperature, respectively. Additionally, nonharmonic and anharmonic entropy terms were calculated as $S_{vib,nh} = S_{vib} - S_{vib,h}$, and $S_{vib,ah} = S_{vib} - S_{vib,h} - S_{vib,d}$, respectively. As can be observed from Fig.~\ref{entropy_Al}, at low temperatures ($T$ = 450\,K ${< 0.5T_m}$, where ${T_m\cong933}$\,K is the melting temperature), harmonic and dilational entropy account well for the total entropy of the system. However, at higher temperature, the anharmonic contribution to the entropy becomes comparable to the dilational contribution. Our results agree remarkably well with entropy calculations provided by Kresch \cite{Kresch}. Small differences in our calculation are primarily due to the higher statistical quality of the phonon DOS reported in Ref.~\citenum{Tang2010} compared to Ref.~\citenum{Kresch}.

\begin{figure}
\begin{center}
\includegraphics[trim=29.0cm 0.5cm 5cm 0.5cm, clip=true, width=0.45\textwidth]{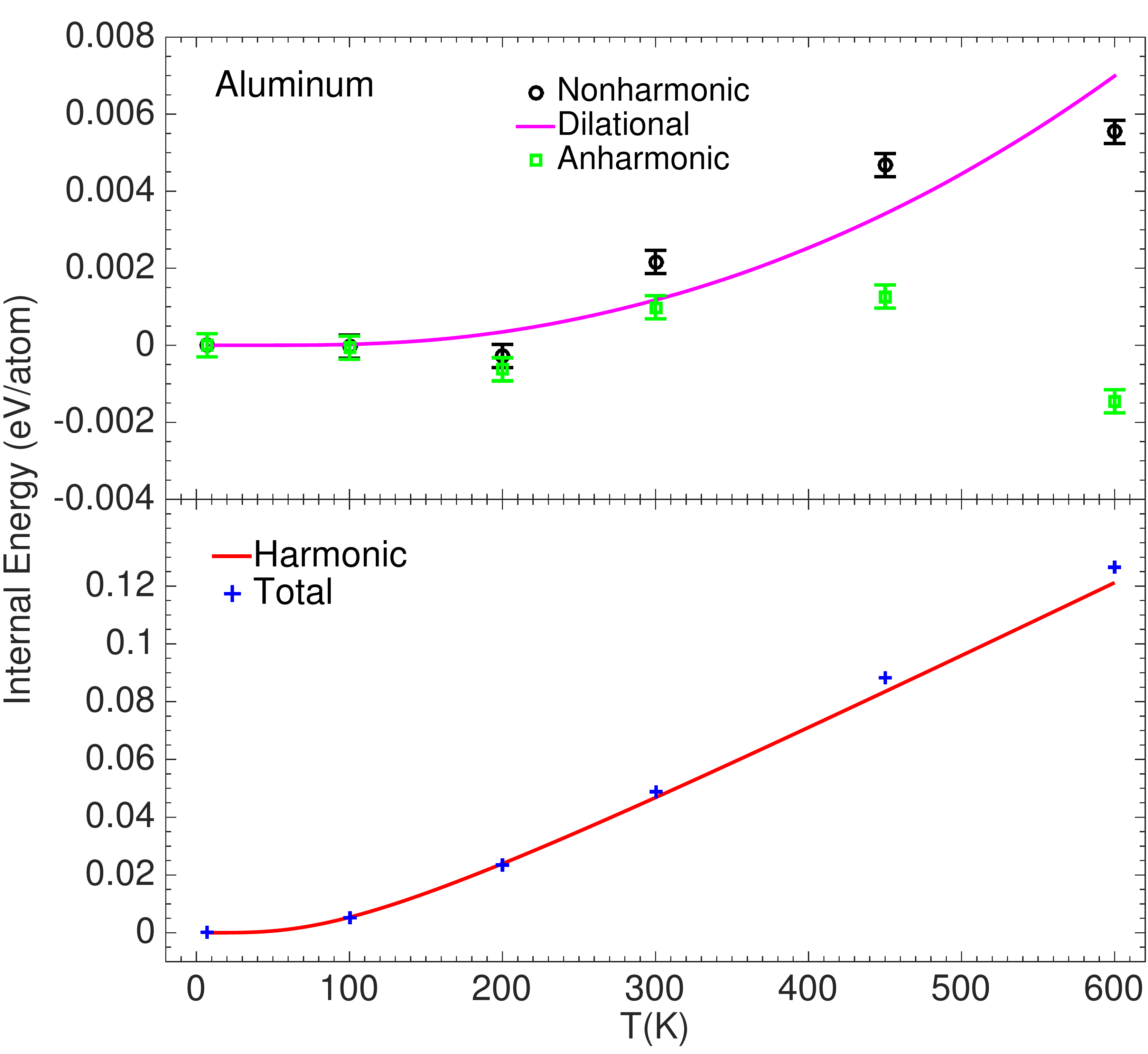}% Here is how to import EPS art
\caption{\label{internal_E_Al}Variation of harmonic, nonharmonic, total, dilational, and anharmonic internal energy $E = E(T) - E(T=0\,K)$ of aluminum crystal with temperature.}
\end{center}
\end{figure}

Figure~\ref{internal_E_Al} shows the harmonic, nonharmonic, total, dilational, and anharmonic internal energy terms for aluminum. The harmonic $E_h$, nonharmonic $E_{nh}$, dilational $E_d$, and anharmonic $E_{ah}$ internal energy were calculated using Eq.~\eqref{ENS1}, $E_{nh} = E - E_h$, Eq.~\eqref{ENS16}, and $E_{ah} = E - E_h - E_d$, respectively, where 
\begin{align}\label{ENS16}
E_d = \int\limits_0^T~(C_p-C_v)~dT'= \int\limits_0^T~\frac{T'\alpha_{vib}(T')^2V(T')}{\chi(T')}~dT'
\end{align}
Again, we can observe that the total internal energy is consistently higher than the harmonic internal energy of the lattice, and the anharmonic contribution to the internal energy becomes significant for ${T > 0.5T_m}$. We also note that the anharmonic vibrational entropy and anharmonic internal energy both become negative around 500\,K, where the dilational contribution exceeds the difference between total and harmonic component. 

\begin{figure}
\begin{center}
\includegraphics[trim=29.0cm 0.5cm 5cm 0.5cm, clip=true, width=0.45\textwidth]{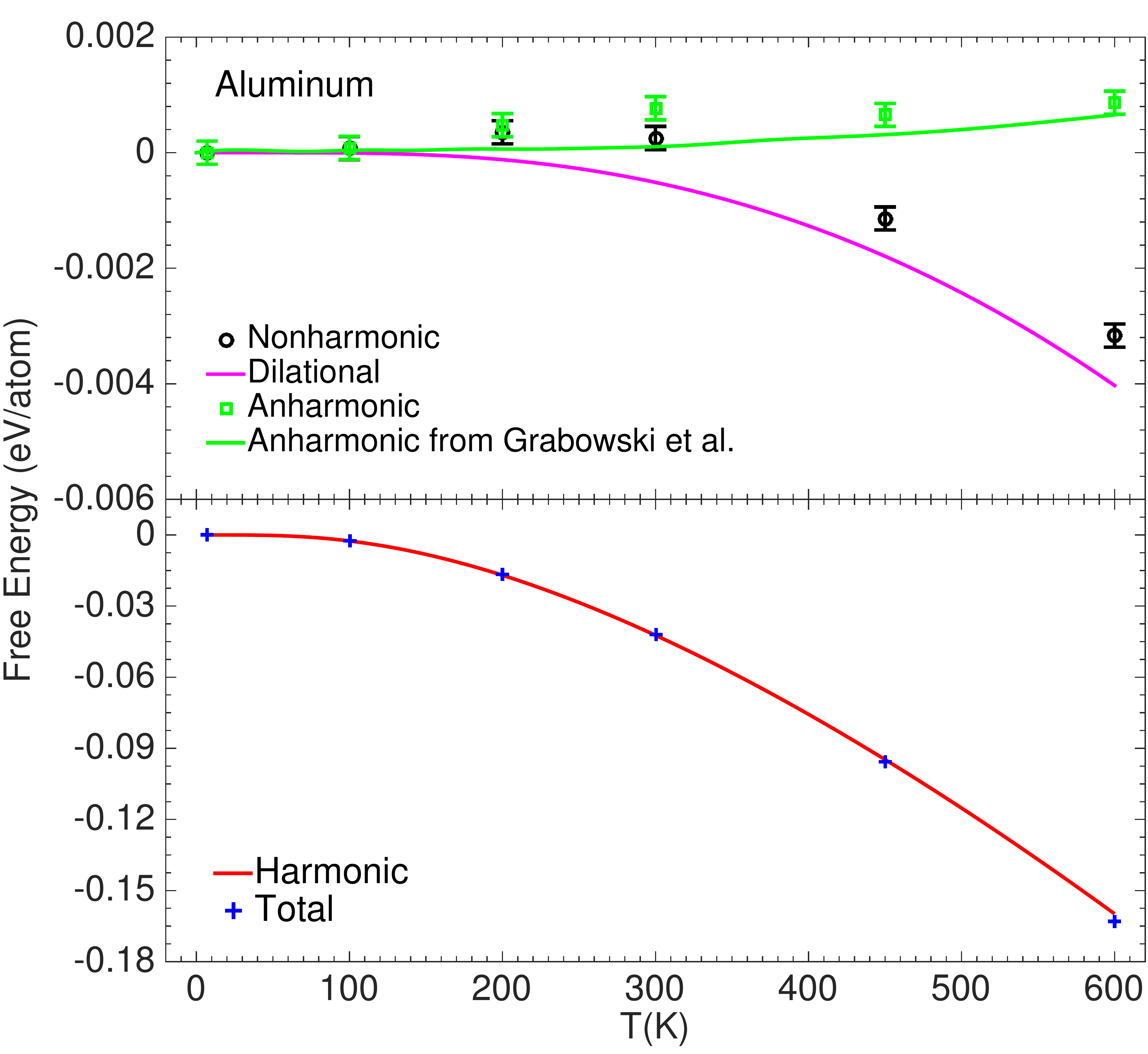}% Here is how to import EPS art
\caption{\label{free-energy_Al}Variation of harmonic, nonharmonic, total, dilational, and anharmonic vibrational free energy $F = F(T) - F(T=0\,K)$ of aluminum crystal with temperature. Anharmonic vibrational free energy is also compared against DFT simulations from Grabowski \emph{et al.}~\cite{Grabowski2009}.}
\end{center}
\end{figure}

Figure~\ref{free-energy_Al} shows the harmonic, nonharmonic, total, dilational, and anharmonic vibrational free energy of aluminum. The harmonic $F_{vib,h}$, nonharmonic $F_{vib,nh}$, total $F_{vib}$, dilational $F_{vib,d}$, and anharmonic $F_{vib,ah}$ internal energy were  calculated using Eq.~\eqref{ENS5} with the same notation as for entropy and internal energy. Due to the increased entropy of the system at higher temperature, the total free energy is suppressed by $\sim$160\,meV at 600\,K. The harmonic free energy accounts for most of this, or $\sim$157\,meV at 600\,K. The dilational free energy, owing to the change in the vibrational frequencies due to the increased lattice volume, contributes approximately $-4$\,meV at 600\,K, with the anharmonic free energy corresponding to the remainder between total and harmonic free energy. Our estimate of the vibrational anharmonic free energy value (0.8$\pm$0.2\,meV/atom at 600\,K) is quantitatively in excellent agreement with the most comprehensive quasi-harmonic and anharmonic {\it ab initio} studies of Grabowski \emph{et al.}~\cite{Grabowski2007,Grabowski2009}.

\subsection{FeSi}

\begin{figure}
\begin{center}
\includegraphics[trim=29.0cm 0.5cm 5cm 0.5cm, clip=true, width=0.45\textwidth]{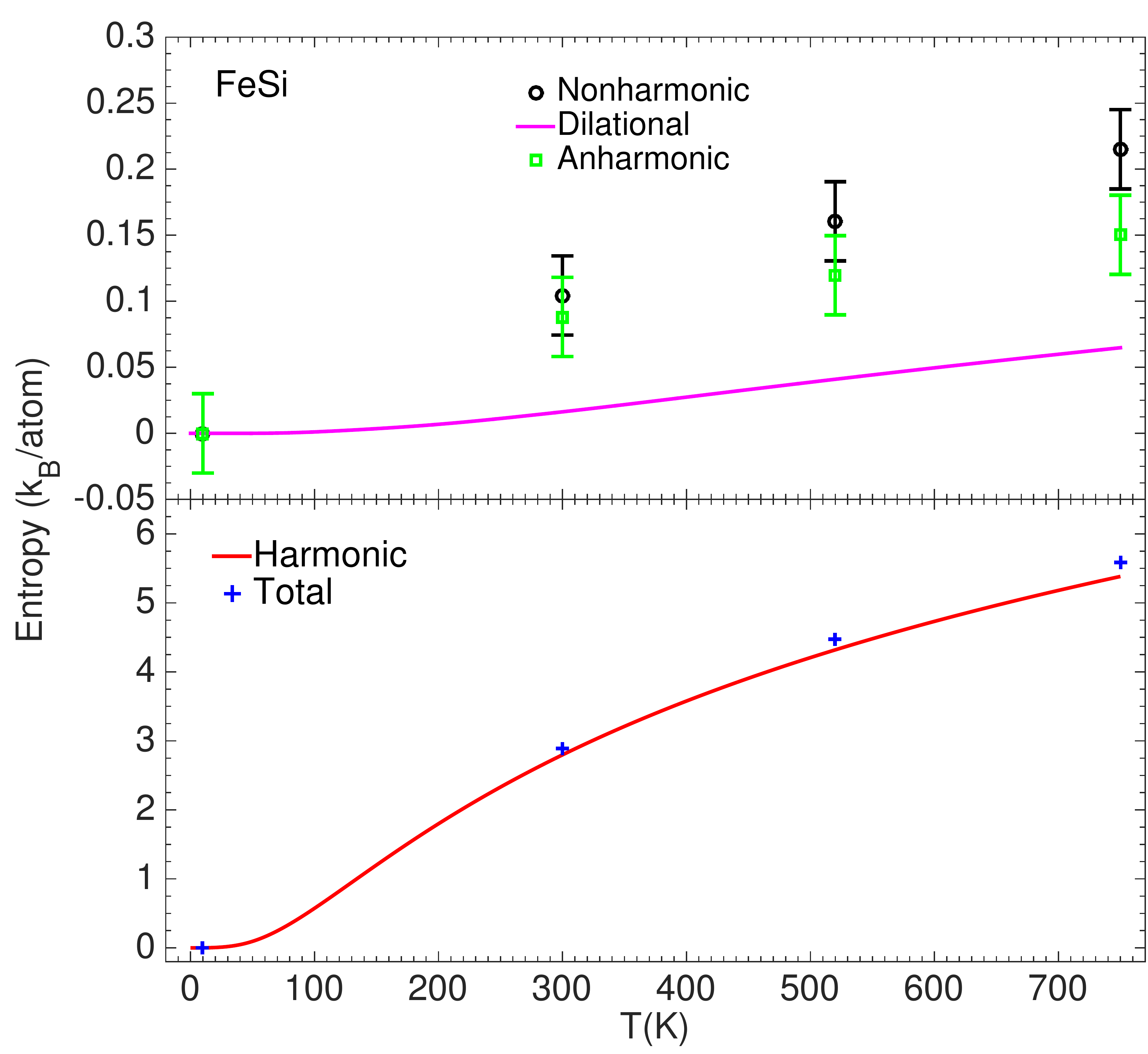}% Here is how to import EPS art
\caption{\label{entropy_FeSi}Variation of harmonic, nonharmonic, total, dilational, and anharmonic vibrational entropy $S = S(T) - S(T=0\,K)$ of ${\rm FeSi}$ crystal with temperature. The error in entropy values are estimated from the error bar in phonon density of states.}
\end{center}
\end{figure}

\begin{figure}
\begin{center}
\includegraphics[trim=29.0cm 0.5cm 5cm 0.5cm, clip=true, width=0.45\textwidth]{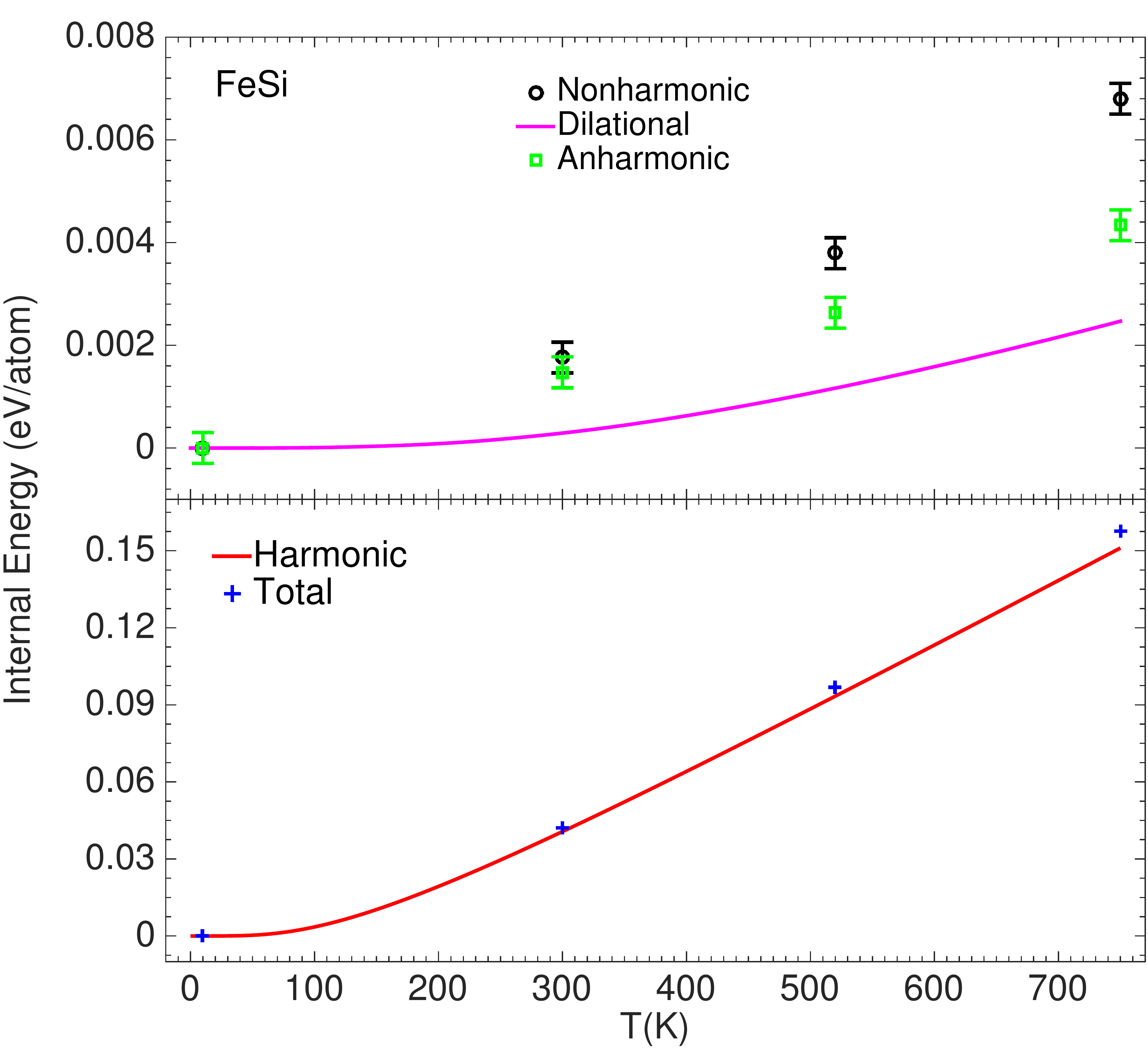}% Here is how to import EPS art
\caption{\label{internal_E_FeSi}Variation of harmonic, nonharmonic, total, dilational, and anharmonic internal energy $E = E(T) - E(T=0\,K)$ of ${\rm FeSi}$ crystal with temperature.}
\end{center}
\end{figure}

\begin{figure}
\begin{center}
\includegraphics[trim=29.0cm 0.5cm 5cm 0.5cm, clip=true, width=0.45\textwidth]{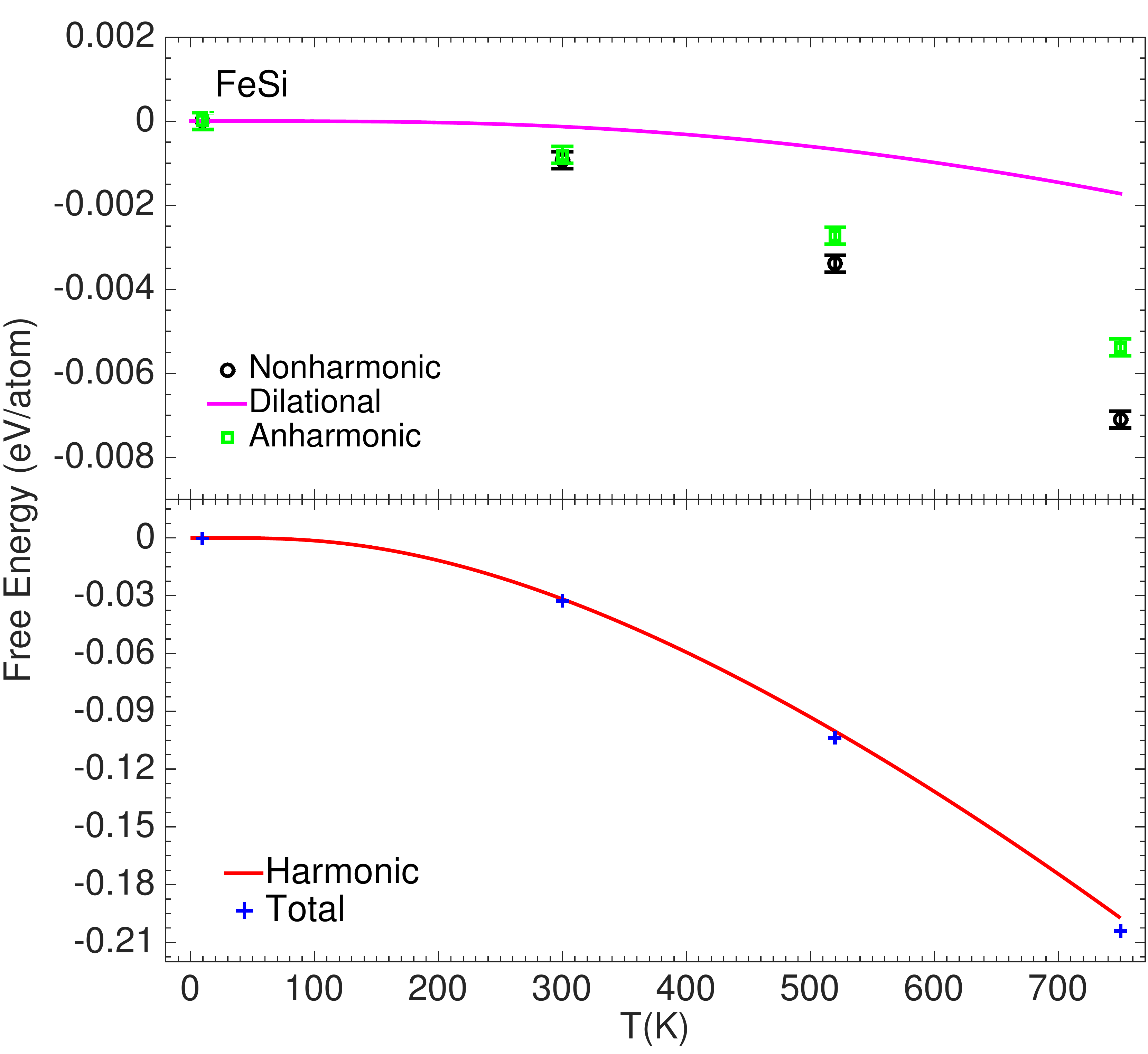}% Here is how to import EPS art
\caption{\label{free-energy_FeSi}Variation of harmonic, nonharmonic, total, dilational, and anharmonic vibrational free energy $F = F(T) - F(T=0\,K)$ of ${\rm FeSi}$ crystal with temperature.}
\end{center}
\end{figure}

Iron mono-silicide is a narrow gap semiconductor crystallizing in the B20 cubic structure, which undergoes a diffuse metal-insulator transition around 200\,K on heating. The strong renormalization of the electronic structure by thermal disorder was shown to cause an anomalous phonon softening, revealed with $T$-dependent INS measurements \cite{Delaire_2011_PNAS, Delaire2013}. Following the same derivation as in the case of aluminum described above, we quantify the various components of the vibrational entropy, internal energy and free energy. To evaluate the expression in Eq.~\eqref{ENS14}, we used experimental volumetric thermal expansion data from Vocadlo \emph{et al.} \cite{Vocadlo2002}, and temperature dependent compressibility values from Petrova \emph{et al.} \cite{Petrova2010} and Povzner \emph{et al.} \cite{Povzner2015}. The vibrational thermal expansion coefficient $\alpha_{vib}$ was calculated by subtracting the electronic thermal expansion coefficient \cite{Mandurus1994, Mandurus1995} from the total thermal expansion coefficient (the role of vacancies is thought to be negligible in FeSi in the temperature range considered here). 

The results for harmonic, nonharmonic, total, dilational, and anharmonic vibrational entropy, internal energy, and free energy are shown in Fig.~\ref{entropy_FeSi},~\ref{internal_E_FeSi}, and ~\ref{free-energy_FeSi}, respectively. The overall behavior of FeSi crystal is similar to that observed in aluminum; although with one striking difference. As can be observed from Fig.~\ref{entropy_FeSi}--\ref{free-energy_FeSi}, the dilational contribution of entropy, internal energy and free energy, originating from the thermal expansion of the crystal, decrease the difference between total and harmonic component. Since the remaining originates from the explicit temperature dependence of vibration frequencies, i.e., anharmonic effects, the difference cannot be completely cancelled out by the thermal expansion of the crystal alone. Consequently, the anharmonic contribution to the entropy and internal energy is positive, while it is negative in the free energy. This is opposite to the behavior seen in aluminum. This positive contribution to the anharmonic free energy in Al implies that the thermal expansion due to the anharmonic effects is negative. Indeed, in their study, Grabowski \emph{et al.}~\cite{Grabowski2009} did calculate a negative contribution to the thermal expansion of aluminum owing to anharmonicity. We also note that the strong renormalization of the electronic structure in FeSi leads to a larger ``anharmonic'' behavior (following terminology introduced above), but that this arises primarily through the intrinsic $T$ dependence of force-constants through the coupling with the electronic structure (gradual metallization of the system, with $T$-dependent interatomic potentials, although remaining quadratic) \cite{Delaire_2011_PNAS, Hellman2013}. The anharmonic contributions to vibrational entropy, internal energy, and free energy are estimated to be 0.15$\pm$0.03\,$k_{\rm B}$, 4.3$\pm$0.3\,meV, and $-5.4\pm$0.2\,meV per atom, respectively at 750\,K, significantly larger in magnitude than that of  aluminum.

In essence, both dilational and anharmonic components are essential to accurately describe high-temperature thermodynamics. In aluminum, the vibrational entropy from the QHA is only accurate within 10\% for low to medium temperatures (below 450\,K), while anharmonic contributions become significant at higher temperature. FeSi displays a more strongly anharmonic behavior with anharmonic effects contributing, in particular, to increasing its thermal expansion.

\section{Conclusion}\label{conclusion}

We derived easily-assessed expressions relating the internal energy, vibrational free energy, volumetric thermal expansion coefficient, with experimental input from thermal expansion and phonon DOS measurements. This approach facilitates a quantitative understanding of how competing thermodynamic forces, including anharmonic effects, influence the crystalline lattices at high temperature. The anharmonic contribution to entropy, internal energy, and free energy for aluminum have been calculated to be -0.04$\pm$0.03\,$k_{\rm B}$, -1.5$\pm$0.3\,meV, and 0.8$\pm$0.2\,meV per atom, respectively at 600\,K, in excellent agreement with {\it ab initio} studies delineated in the literature. Similarly, the anharmonic contribution to entropy, internal energy, and free energy for ${\rm FeSi}$ have been calculated to be 0.15$\pm$0.03\,$k_{\rm B}$, 4.3$\pm$0.3\,meV, and -5.4$\pm$0.2\,meV per atom, respectively at 750\,K, significantly larger in magnitude than that of aluminum. 

\section{Acknowledgements}

Work at ORNL was supported through the Office of Science Early Career Research Program (PI O.D.). Work at UB was supported through MCEER.

%\bibliography{TENS}
%\bibliography{references1}
%\bibliography{apssamp}% Produces the bibliography via BibTeX.
%

\end{document}